\documentclass[aps,pre,reprint,groupedaddress,showpacs]{revtex4-1}
\usepackage[english]{babel}
\usepackage[latin2]{inputenc}
\usepackage{graphicx}
\usepackage{amsmath}
\usepackage{rotating}

\begin{document}
\author{M. Majka}
\email{maciej.majka@uj.edu.pl}
\affiliation{Marian Smoluchowski Institute of Physics, Jagiellonian University, Reymonta 4, 30-059 Kraków, Poland}
\author{P. F. G\'{o}ra}
\affiliation{Marian Smoluchowski Institute of Physics, Jagiellonian University, Reymonta 4, 30-059 Kraków, Poland}

\title{Polymer unfolding and motion synchronization induced by spatially correlated noise}

\begin{abstract}
The problem of a spatially correlated noise affecting a complex system is studied in this paper. We present a comprehensive analysis of a 2D model polymer chain, driven by the spatially correlated Gaussian noise, for which we have varied the amplitude and the correlation length. The chain model is based on a bead-spring approach, enriched with a global Lennard-Jones potential and angular interactions. We show that spatial correlations in the noise inhibit the chain geometry dynamics, enhancing the preservation of the polymer shape. This is supported by the analysis of correlation functions of both the module length and angles between neighboring modules, which have been measured for the noise amplitude ranging over 3 orders of magnitude. Moreover, we have observed the correlation length dependent beads motion synchronization, and the spontaneous polymer unfolding, resulting from an interplay between chain potentials and the spatially structured noise.
\end{abstract}

\maketitle

%-----------------INTRODUCTION-----------------
\section{Introduction}
The understanding of diffusion in complex media is crucial for both the modeling of conformation transitions in biomolecules and the intracellular transport. It is also well known that various systems organize spontaneously in a response to random forcing \cite{bib:sagues} and that the introduction of temporal correlations into the noise can lead to synchronization effects \cite{bib:longa}. A well-established framework to simulate these phenomena is provided by Langevin equations, which introduce the concept of stochastic force mimicking the molecular collisions \cite{bib:gardiner}. An important advance in this formalism has been the introduction of Generalized Langevin Equation (GLE), which reproduce the anomalous diffusion thanks to the time-correlated stochastic force and the corresponding integral memory kernel, which represents the friction \cite{bib:kou}. Recently, S. C. Kou in \cite{bib:kou} has derived the GLE from a microscopic model of a particle coupled to a large number of oscillators, thus showing that the particle-environment interaction is essential for the occurrence of temporal correlations in thermal noise. However, it is remarkable that this theory explains solely the temporal aspect of diffusion, while little work has been done to understand its spatial counterpart. This has led us to investigate the problem of a spatially correlated noise affecting a complex system.

The collective media behavior which is random but characterized by a certain correlation length $\lambda$ occurs at the length-scale of micrometers in the context of hydrodynamic interactions, e.g. in colloid sedimentation \cite{bib:sed1,bib:sed2} or in the study of the active particle motion \cite{bib:swimmers}. However, the spatial correlations at the lower length-scale play a fundamental role in the theory of phase transitions \cite{bib:binney} among which the liquid-glass transition is of special interest. During this transition, the particles suffer a dramatic drop of mobility without the emergence of structural ordering \cite{bib:mosayebi}. This phenomenon has been intensively researched on for past two decades, and, according to extensive simulations \cite{bib:donati, bib:doliwa}, it is characterized by the occurrence of the spatial correlations in particles motion \cite{bib:doliwa}, which is recognized as the formation of the different-sized clusters \cite{bib:donati, bib:mitus}. Choosing a single moment in time, one could interpret these clusters as a source of a disturbance which is random at the large length scale ($\gg\lambda$), but ordered at the length scale of $\lambda$. Figure \ref{pic:vec_field} illustrates this idea. The temporal evolution of this system is still indeterministic, as it 'randomly reorders'. We propose that this behavior could be imitated by the spatially correlated noise, which is affecting a subsystem, in our case, a model 2D polymeric chain. 

We have simulated the chain based on the bead-spring approach under the forcing of Spatially Correlated Gaussian Noise (SCGN) for which we have varied the correlation length and the amplitude. Our previous findings regarding the stiffening of the chain under the SCGN, shown with the aid of the reduced dynamics, have been published in \cite{bib:mm}. However, our further investigation into this system, which involves the extension of the parameters' range and the measurements of chain characteristics, has revealed several new effects, namely: beads motion synchronization, increased time correlation of both module length and angles between modules, the inhibition of the average module length growth and, most notably, the chain unfolding induced by the increased correlation length.

Our simulations are related to the actual physical situation by the choice of $\lambda$. Unfortunately, currently there are few experimentally accessible quantities that describe the collective molecular behavior in the vicinity of glass transition and can be measured for the variety of temperatures \cite{bib:calorimetry}. One of these parameters is the number of cooperatively rearranging molecules \cite{bib:impendspec}, which has been reported to rise from 1 in liquid phase to approximately 10 in the glass phase \cite{bib:impendspec}. Additionally, these results are qualitatively similar for the different chemical compounds \cite{bib:impendspec}. On the other hand, the direct measurements of the correlation length are scarce and limited to specific experimental setup, as e.g. in \cite{bib:confocal} which reports $\lambda$ to be of the order of 2-4 molecule's diameters. These measurements suggest that $\lambda$ covering up to 5 chain nodes is physically meaningful.

The paper has following structure: in section \ref{sec:scgn} the methods of the SCGN generation are introduced, in section \ref{sec:equations} we propose the equations of motion and the correlation function, in section \ref{sec:pol_model} we present our polymer model, section \ref{sec:sim_methods} briefly discusses  simulation methods, sections from \ref{sec:synchronization} to \ref{sec:unfolding} present the results regarding each effect with an interpretation, in section \ref{sec:summary} we summarize our findings. 

%--------------SPATIALLY CORRELATED NOISE---------------------
\section{Multiple correlated Gaussian variables} \label{sec:scgn}
The generation of multiple correlated Gaussian variables is a central problem in the simulation of SCGN driven systems, therefore we shall outline here the basic algorithm.

Lets assume that we have two real vectors of random, zero-mean Gaussian variables, namely $\vec{ \xi}^T=(\xi_1,\dots,\xi_N)$ and $\vec{ \eta} ^T=(\eta_1,\dots,\eta_N)$, which components satisfy the following correlation relations:
\begin{gather}
\left<\xi_i\xi_j\right>=S_{ij} \\
\left<\eta_i\eta_j\right>=\delta_{ij}
\end{gather}
Here, $\delta_{ij}$ denotes the Kronecker delta, and $S_{ij}$ are elements of  the correlation matrix, defined as:
\begin{equation}
\left<\vec{\xi}\vec{\xi}^T\right>=\hat{S}
\end{equation}
The matrix $\hat{S}$ is symmetric and positively definite \cite{bib:kubo}, so it is suitable for Cholesky decomposition \cite{bib:golub}, which factorizes $\hat{S}$ into a lower triangular matrix $\hat{L}$ and its transposition:
\begin{equation}
\hat{S}=\hat{L}\hat{L}^T
\end{equation}
The vector of correlated variables $\vec{\xi}$ is related to the uncorrelated vector $\vec{\eta}$ via a linear transformation \cite{bib:wieczorkowski}:
\begin{equation}
\vec{\xi}=\hat{L}\vec{\eta} \label{eq:xi}
\end{equation}
This means that, given a correlation matrix, one can generate the correlated Gaussian vector $\vec{\xi}$ simply by sampling $N$ times the normal distribution to obtain the components of $\vec{\eta}$ and then performing the transformation \eqref{eq:xi}. 

%------------GENERAL METHODOLOGY----------------------------------
\section{Equations of motion and correlation function} \label{sec:equations}
Our system is equivalent to an ordered set of $N$ interacting material points on a plane, enumerated by the index $i$. The position of $i$-th point (or bead, as we will refer to it further) is $\vec{r_i}^T=(x_i,y_i)$. In order to simulate the trajectory $\{\vec{r}_i(t)\}_N$ of the whole system, we have to solve numerically a set of $2N$ stochastic equations of motion:
\begin{equation} \label{eq:main}
\left\{
\begin{aligned}
m\ddot{x_i}+\gamma\dot x_i+\partial_{x_i}U=\xi_{x}(\vec{r_i}) \\
m\ddot{y_i}+\gamma\dot y_i+\partial_{y_i}U=\xi_{y}(\vec{r_i})
\end{aligned}
\right.
\end{equation}
Here, $U$ is the potential energy of the system, which we will discuss in detail in the next section. $\vec{\xi}(\vec{r_i})^T=\left(\xi_{x}(\vec{r_i}),\xi_{y}(\vec{r_i})\right)$  is the two dimensional SCGN, $m$ is a bead mass and $\gamma$ is a friction constant. 
In an absence of a more relevant theory, we have applied the simplest friction model, and chosen $\gamma$ to be constant. The differential equations \eqref{eq:main} are in principle of the second order, which we preserve for generality, but in the course of our simulations we have overdamped the system by choosing $\gamma$ to be large enough.

We assume that the correlation function $S_{ij}$ of stochastic forces acting on beads $i$ and $j$ should depend only on a relative distance between these beads, which is $r_{ij}=|\vec{r_i}-\vec{r_j}|$. Additionally, we also assume that there are no cross correlations between $x$ and $y$ components, which allows us to reduce the correlation relations to the form:
\begin{equation} \label{eq:corr_simp}
\begin{gathered}
\left<\xi_{x}(\vec{r_i})\xi_{x}(\vec{r_j})\right>=\left<\xi_{y}(\vec{r_i})\xi_{y}(\vec{r_j})\right>=S(r_{ij}) \\
\left<\xi_{x}(\vec{r_i})\xi_{y}(\vec{r_j})\right>=0
\end{gathered}
\end{equation}

It should be emphasized that the correlation matrix $\hat S$ is a dynamical object, and evolves in $t$ as the relative distances $r_{ij}(t)$ do. The conditions \eqref{eq:corr_simp} suggest the following procedure to integrate equations \eqref{eq:main}: once all bead's positions $\{\vec{r_i}(t)\}_N$ at some moment $t$ are determined, we can calculate the $N\times N$ correlation matrix and its Cholesky decomposition $\hat{L}$; next, according to \eqref{eq:xi}, we shall use $\hat{L}$ and two different $\vec{\eta}$ to obtain $\{\xi_x(\vec{r_i})\}_N$ and $\{\xi_y(\vec{r_i})\}_N$. Finally, we can use them to perform an integration step, which gives $\{\vec{r_i}(t+\Delta t)\}_N$. The repeated Cholesky decompositions are the most computationally expansive part of our simulations, as the computational complexity of this decomposition is $O(N^3)$\cite{bib:golub}. 

Along with conditions \eqref{eq:corr_simp}, we assume that the correlation function $S_{ij}$ is characterized by the correlation length $\lambda$, and it reproduces the standard Brownian diffusion for $\lambda \to 0$ \cite{bib:kampen}, so:
\begin{equation}
S\left(r_i(t),r_j(t')\right)\overset{\lambda\to 0}{=}\frac{2kT\gamma}{m}\delta(t-t') \label{eq:brown}
\end{equation}
In the above formula $k$ denotes Boltzmann constant and $T$ is a temperature. Taking into account \eqref{eq:corr_simp} and \eqref{eq:brown}, we chose the exponentially decaying spatial correlation function, which resembles the displacement correlation function from \cite{bib:doliwa} and \cite{bib:confocal}. We also neglect the temporal correlations, as we are interested in the effects of the purely spatially structured noise. Finally, the spatio-temporal correlation function reads:
\begin{equation}
S\left(r_i(t),r_j(t')\right)=\sigma\frac{\gamma}{m}e^{-\frac{|\vec{r}_i-\vec{r}_j|}{\lambda}}\delta(t-t') \label{eq:corr_fun}
\end{equation}
$\sigma=2kT$ denotes the noise amplitude and we will refer to it as temperature, as it is proportional to the actual physical temperature.

\begin{figure}
\includegraphics[width=0.5\textwidth]{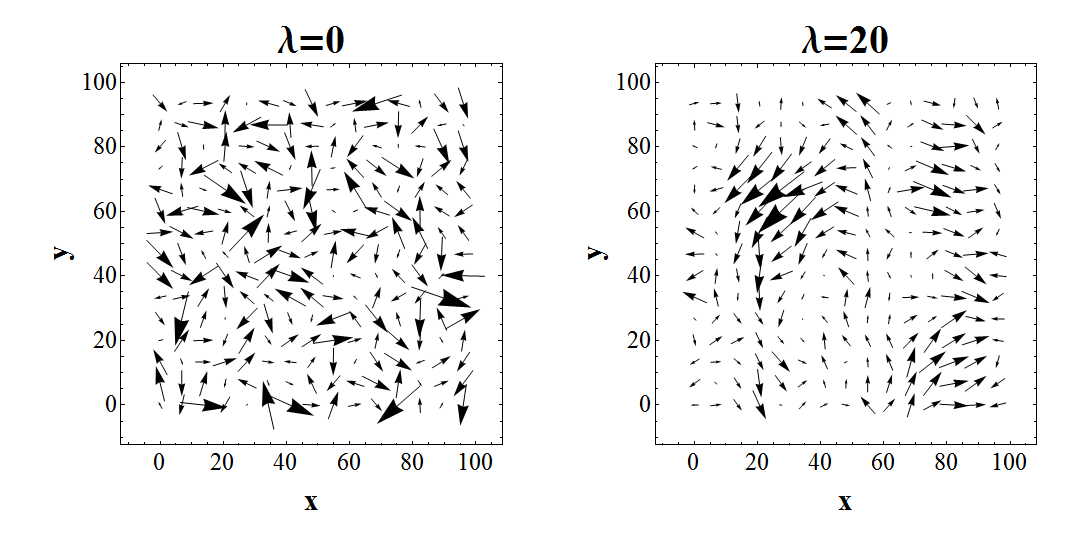}
\caption{A spatially correlated random vectors with correlation length $\lambda$, generated on a regular network. For $\lambda=0$ the pattern is entirely random, but for $\lambda=20$ the ordered clusters can be noticed.\label{pic:vec_field}}
\end{figure}

In order to illustrate how the spatial correlations affect the noise pattern, we have applied \eqref{eq:xi} and \eqref{eq:corr_fun} to generate the random vectors on a regular network. A snapshot from this simulation is presented in Figure \ref{pic:vec_field}. One can easily notice clusters of correlated vectors, however, this pattern changes dramatically for every new generation. 
%figure
%---------------------POLYMER CHAIN MODEL------------------------
\section{The model of polymer chain} \label{sec:pol_model}
The polymeric chain is an archetype of many biomolecules, thus we have chosen it as a test-object for our simulation. Our model is based on the bead-spring approach, in which $i$ and $i+1$ beads interact with a harmonic potential:
\begin{equation}
U_{R}=\sum_{i=1}^{N-1}\frac{1}{2}k_1(|\vec{r}_{i+1}-\vec{r}_{i}|-d_0)^2
\end{equation}
Every bead is also the source of the Lennard-Jones type interaction, which provides excluded volume effect and an interaction between the distant tails of the chain:
\begin{equation}
U_{LJ}=\sum_{i,j}^{N}\epsilon\left(\frac{\sigma^{12}_{LJ}}{|\vec{r}_i-\vec{r}_j|^{12}}-\frac{\sigma^6_{LJ}}{|\vec{r}_i-\vec{r}_j|^6}\right)
\end{equation}
Finally, we introduce a harmonic interaction between beads $i$ and $i+2$, which resembles angular interactions:
\begin{equation}
U_{\psi}=\sum_{i=1}^{N-2}\frac{1}{2}k_2(|\vec{r}_{i+2}-\vec{r}_{i}|-l_0)^2
\end{equation}
The total potential energy $U$ is equal to:
\begin{equation}
U=U_R+U_\psi+U_{LJ} \label{eq:potential}
\end{equation}

For $\epsilon=0$ (no $U_{LJ}$ contribution) the potential energy is minimized when beads' positions satisfy:
\begin{equation} \label{eq:minima}
\left\{
\begin{aligned}
|\vec{r}_{i+1}-\vec{r}_{i}|&=d_0 \\
|\vec{r}_{i+2}-\vec{r}_{i}|&=l_0
\end{aligned}
\right.
\end{equation}
In this case, all of minimum energy conformations are equienergetic. In fact, unless $l_0>2d_0$, once $\vec{r}_{1}$ and $\vec{r}_{2}$ are chosen to satisfy $|\vec{r}_{2}-\vec{r}_{1}|=d_0$, the 3rd bead can be positioned in two ways, so the relation $|\vec{r}_{3}-\vec{r}_{1}|=l_0$ is also fulfilled. Successively applying the conditions \eqref{eq:minima} to following beads, one can built numerous minimum energy geometries. When $U_{LJ}\neq0$ the energetic structure of the chain become more complex, but if $d_0>\sigma_{LJ}$ and $\epsilon\simeq k_1$, the Lennard-Jones contribution becomes a perturbation. However, the $U_{LJ}$ influence makes the structures no longer equienergetic. 

\begin{figure}
\includegraphics[width=0.4\textwidth]{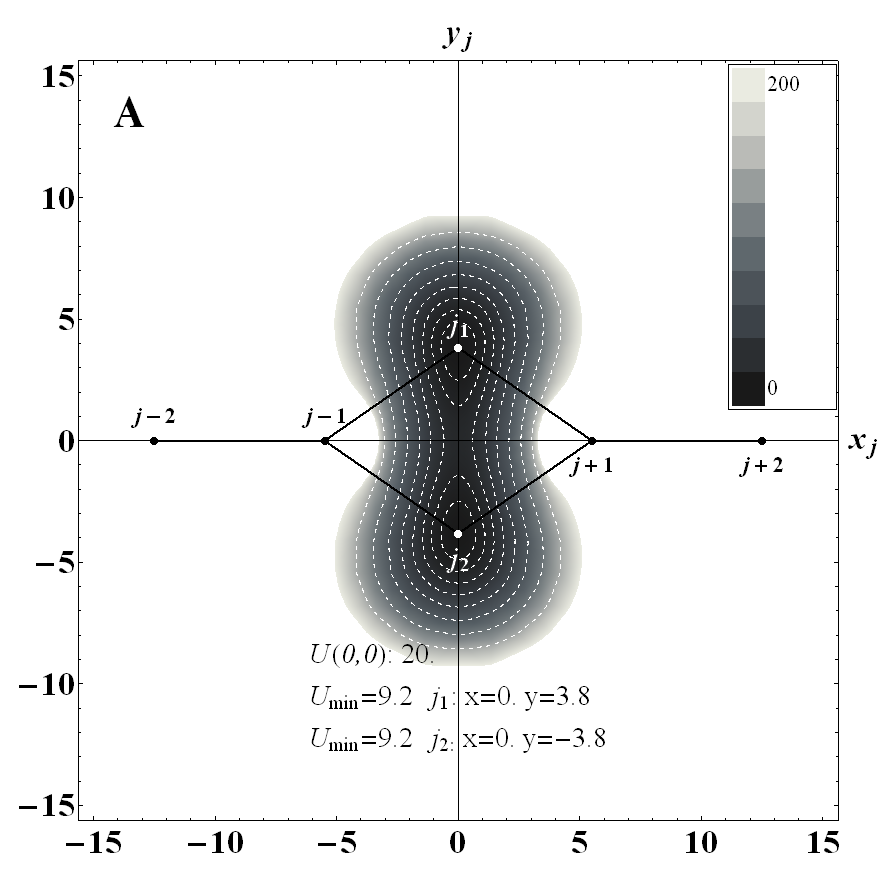}
\includegraphics[width=0.4\textwidth]{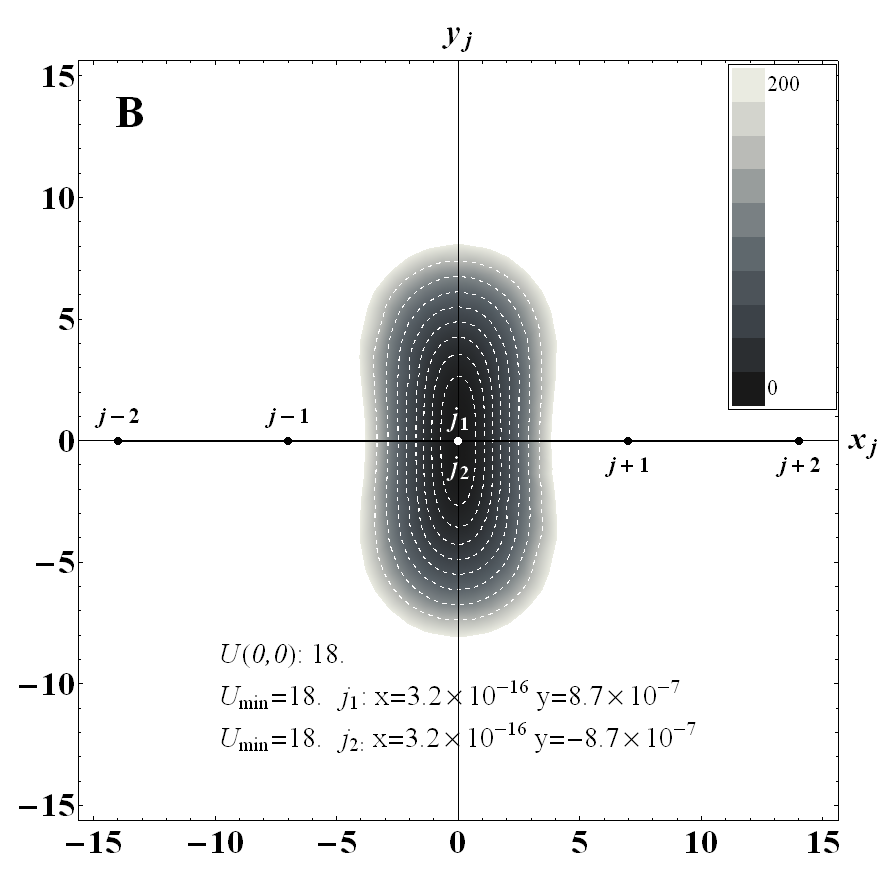}
\caption{The energetic landscape for a single bead interacting with its 4 nearest neighbors. Potential parameters are chosen according to Table \ref{tab:parameters}. A) The distance $|\vec{r}_{j+1}-\vec{r}_{j-1}|=l_0$,  $j_1$ and $j_2$ enumerate two possible positions of $j$-th bead that minimize the potential energy. $U(0,0)$ is the height of energy barrier and $U_{min}$ is the depth of the minimum. B) $|\vec{r}_{j+1}-\vec{r}_{j-1}|=1.25l_0$, $j_1$ and $j_2$ merge into a single minimum as the energy barrier disappears. \label{pic:landscape}}
\end{figure}

\begin{table}
\begin{center}
\begin{tabular}{c c c c c c c c c }
\hline
$N$ & $ k_1$ & $d_0$ & $ k_2$ &$ l_0$ & $\epsilon$ & $\sigma_{LJ}$ &$ \gamma$ & $m$ \\
\hline
128 & 7 & 7 & 2 & 11 & 1 & 3 & 20 & 1 \\
\hline
\end{tabular}
\caption{The parameters of the system chosen for simulation. }\label{tab:parameters}
\end{center} 
\end{table}
When the chain's energy is not minimized, the dynamical topography of the potential energy surface depends on both potentials' parameters and the local geometry of the chain. An effective way to represent snapshots of this energy landscape for a single bead is to take into account its 4 nearest neighbors. An example of such a landscape is reproduced in Figure \ref{pic:landscape}A. We have chosen the values of potential energy parameters (Table \ref{tab:parameters}) such that the double minimum structure is distinct and holds for the wide range of local conformations. However, this structure is extremely sensitive to a single parameter, which is the distance $l_j=|\vec{r}_{j+1}-\vec{r}_{j-1}|$. Whenever $l_j>2d_0$, the two minima tend to merge rapidly into a single one, positioned in-line with the beads $j-1$ and $j+1$. This is reproduced in Figure \ref{pic:landscape}B. This fact significantly affects the high temperature dynamics of the chain, as is show in section \ref{sec:unfolding}.
%figure

%-----------------------SIMULATION METHODS----------------------
\section{Simulation}\label{sec:sim_methods}
Applying the classical Runge-Kutta method modified for stochastic differential equations \cite{bib:kloeden}, we have simulated the system described by equations \eqref{eq:main} with the potential \eqref{eq:potential} and the parameters from Table \ref{tab:parameters}. The number of beads has been set to $N=128$, the bead's mass has been chosen $m=1$ and the friction coefficient $\gamma$ has been set to 20, which overdamped the system. 

In our research, we have explored three regimes of temperature. First, we have varied the noise amplitude $\sigma$ from 0 to 20 units at the interval of 1 unit, and we have increased the correlation length $\lambda$ from 0 to 20 at the interval of 5 units. In the second regime, we have increased $\sigma$ from 25 to 250 at the interval of 25 units, and in the third regime we have explored region from 300 to 1000 units, at the interval of 100 units. For the second and the third regime we have varied $\lambda$ from 0 to 50 at the interval of 10 units. For each pair of $\lambda$ and $\sigma$ we have performed 64 runs, starting from different initial positions. The initial coordinates has been chosen, so the distance between nearest neighbors has been equal to $d_0$, but the angle between modules has been chosen randomly from $\pi/2$ to $3\pi/2$. 

The integration step has been set to 1/128 time unit, and each simulation lasted 2148 time units. The data for the first 100 units has been rejected due to the system thermalization.  If not stated otherwise, the data has been collected once per time unit. We have gathered the data regarding beads synchronization, module length and angle between modules.

%-------------------SYNCHRONIZATION-------------------------------
\section{The beads motion synchronization}\label{sec:synchronization}
The introduction of the spatial correlations into the noise implies that, at the length scale comparable to the correlation length $\lambda$, the stochastic force vectors have similar direction and value. Therefore, one could expect that the motion of beads, which relative distance is lower than $\lambda$, will synchronize. This prediction has been fully confirmed.

As the measure of synchronization at a particular moment $t$, we have chosen the normalized product of two beads' velocities, distanced by $n$ nodes, which has been averaged along the chain:
\begin{equation}
K_n(t)=\frac{1}{(N-n)}\sum_{i=1}^{N-n}\frac{\vec{v}_i\circ\vec{v}_{i+n}}{v_i v_{i+n}}=\left< \cos\theta_{i,i+n} \right>
\end{equation}
Here $\theta_{i,i+n}$ is an angle between velocity vectors of the $i$-th and $i+n$-th bead. For each run, we have gathered $K_n(t)$, which was time-averaged to obtain synchronization factor $K_n$. The maximal value of $K_n=1$ indicates a fully synchronized motion, while $K_n=0$ implies the opposite. 

%figure
\begin{figure*}
\includegraphics[width=\textwidth]{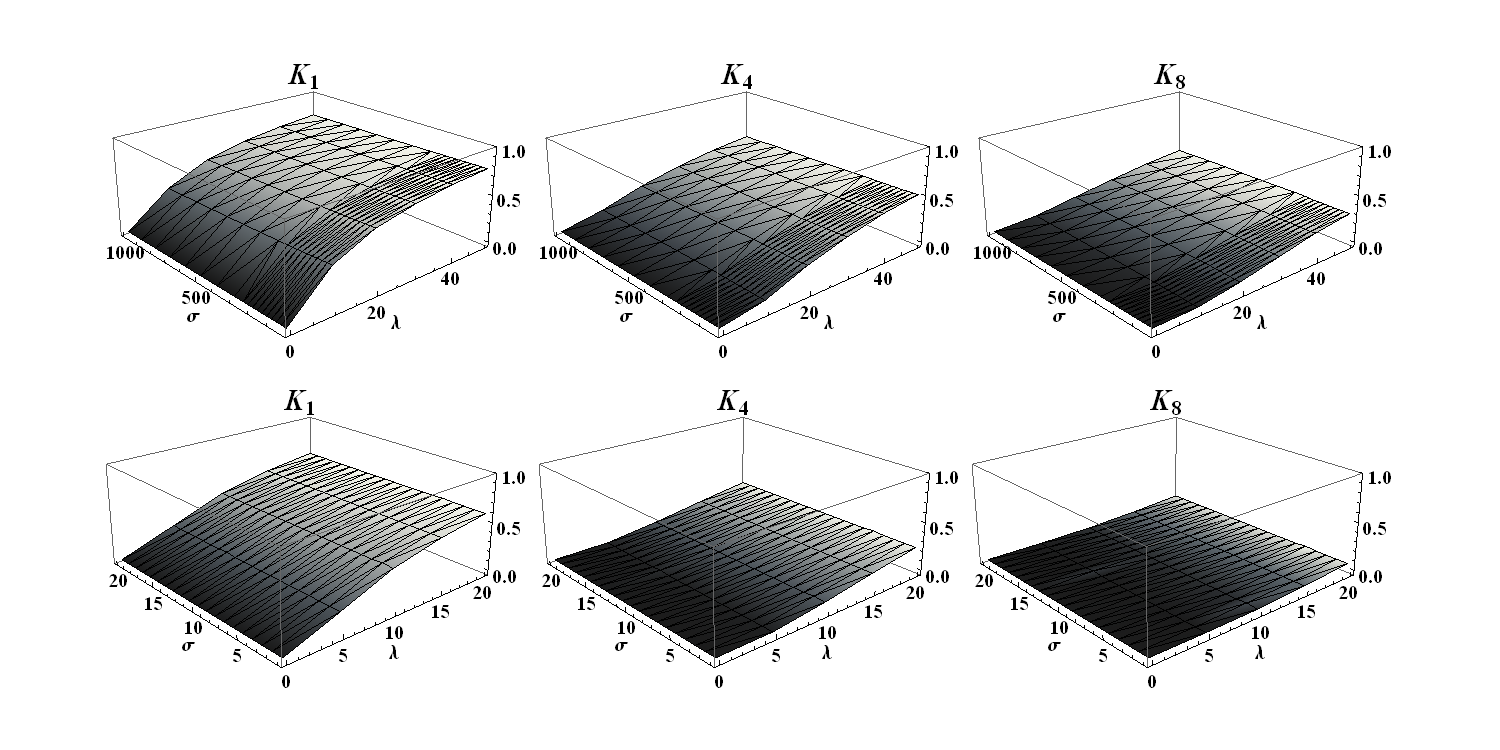}
\caption{The synchronization factor $K_n$ for $n=1,4,8$ as a function of the correlation length $\lambda$ and the noise amplitude $\sigma$. Each column contains the data for the same $n$ in the high temperature regime (upper graph) and the low temperature regime (lower graph).  \label{pic:synch1}}
\end{figure*}
We have gathered the data for $n$ ranging from 1 to 9. A representative sample of our results is show in Figure \ref{pic:synch1}. The rise in the synchronization factor $K_n$ along with increasing $\lambda$ and $\sigma=const$ is evident. Conversely, the level of synchronization is almost constant for $\lambda=const$ and varying amplitude, which is valid even for temperatures below $\sigma=5$. For every $n$, the factor $K_n$ grows from 0 for $\lambda=0$ to the maximal observed value for $\lambda=50$, which is approximately 0.8 for $n=1$ and 0.3 for $n=9$.
%figure
\begin{figure}
\includegraphics[width=0.5\textwidth]{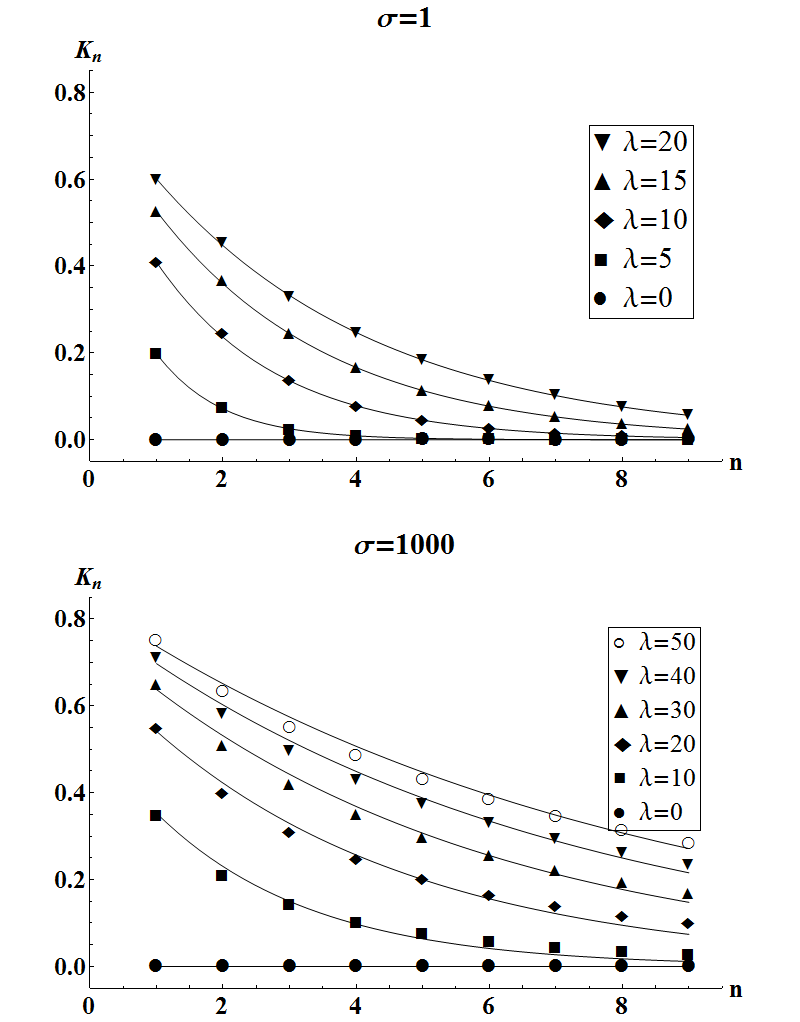}
\caption{The synchronization factor $K_n$ for two extreme values of temperature, as a function of the beads' distance $n$. The data has been fitted with the model: $K_n=A_\lambda e^{-B_\lambda n}$. \label{pic:synch2}}
\end{figure}

A further insight into the synchronization comes from the rearrangement of data, so $K_n$ is represented as a function of $n$ with $\lambda$ and $T$ being parameters. Figure \ref{pic:synch2} shows the qualitative similarity between these data for two extreme temperatures ($\sigma=1$ and $\sigma=1000$). To obtain a quantitative measure of the decrease in synchronization with the rise in $n$, we have fitted our data with the exponential decay model:
\begin{equation}
K_n=A_\lambda e^{-B_\lambda n} \label{eq:decay_model}
\end{equation}
This model proved to be an accurate description of data, as the coefficient of determination $R^2$ exceeded 0.99 for all fits, except those with $\lambda=0$, for which $B_{\lambda=0}$ has no physical meaning. 

\begin{figure}
\includegraphics[width=0.5\textwidth]{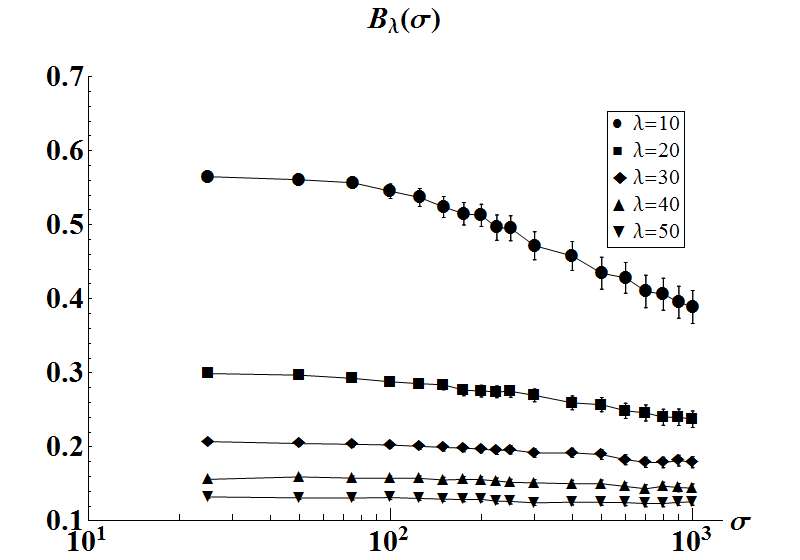}
\caption{The synchronization decay factor $B_\lambda$ as the function of temperature, for $\sigma\ge25$.\label{pic:synch3}}
\end{figure}

In Figure \ref{pic:synch3}, we have juxtaposed the values of $B_\lambda$ for $\sigma\ge25$, at which temperature the behavior of chain is noise-dominated. According to this figure, the value of $B_\lambda$ is mainly determined by $\lambda$ and decreases when the temperature grows by two orders of magnitude. However, for $\lambda\ge30$ this fall is rather insignificant, thus, we conclude that the noise correlation length is the primary factor that influences the effective range of synchronization along the chain.

%figure

%--------------------------------CORRELATIONS----------------------------
\section{Beads motion correlation}\label{sec:correlations}
The other quantities that are also affected by the presence of spatial correlations in noise are: the time correlation of the modules length and the time correlation of the angles between neighboring modules. These two characteristics describe the time evolution of the chain geometry.

By a module we understand two neighboring beads, so the length of the $j$-th module, at certain moment $t$, is defined as:
\begin{equation}
d_j(t)=\left| \vec{r}_j(t)-\vec{r}_{j-1}(t)\right|
\end{equation}
The angle between two neighboring modules is defined by the positions of the three following beads:
\begin{equation}
\psi_j(t)=\angle \left( \vec{r}_{j-1}(t),\vec{r}_{j}(t),\vec{r}_{j+1}(t)\right) \label{eq:psi_j}
\end{equation}
With the beginning at the center of the coordinate system, vectors $\vec{r}_i$ are equivalent to the coordinates on a plane, thus they are applied in the above definition. Additionally, one has to remember that the angle $\psi_j$ is directed and varies from $0^\circ$ to $360^\circ$ (with $180^\circ$ indicating that three beads are exactly in-line), so the angles have to be measured in a unified way along the whole chain, conserving the initial numeration of beads.

The time correlation function of angles has been calculated in a following way:
\begin{equation}
C_\psi(\tau)=\frac{1}{C_\psi}\sum_{k=0}^{T}\sum_{j=2}^{N-1} (\psi_j(t_k+\tau)-\left<\psi\right>)(\psi_j(t_k)-\left<\psi\right>)
\end{equation}
Here, we introduce the additional summation over $j$ due to the fact, that we have $N-2$ angles for a single moment $t$, which allows us to increase statistics and obtain a correlation measure for a whole chain, rather than a single site. The normalization factor $C_\psi$ has been chosen as:
\begin{equation}
C_\psi=\sum_{k=0}^{T}\sum_{j=2}^{N-1} (\psi_j(t_k)-\left<\psi\right>)^2
\end{equation}
which means that $C_\psi(0)=1$. Finally, $\left<\psi\right>$ reads:
\begin{equation}
\left<\psi\right>=\frac{1}{T(N-2)}\sum_{k=0}^{T}\sum_{j=2}^{N-1} \psi_j(t_k)
\end{equation}
In a strict analogy to the angle correlation function $C_\psi(\tau)$, we can define the module length correlation function $C_d(\tau)$:
\begin{equation}
C_d(\tau)=\frac{1}{C_d}\sum_{k=0}^{T}\sum_{j=2}^{N} (d_j(t_k+\tau)-\left<d\right>)(d_j(t_k)-\left<d\right>)
\end{equation}
The $C_d$ and $\left< d\right>$ are defined similarly to their angle counterparts.
%figure
\begin{figure*}
\includegraphics[width=0.9\textwidth]{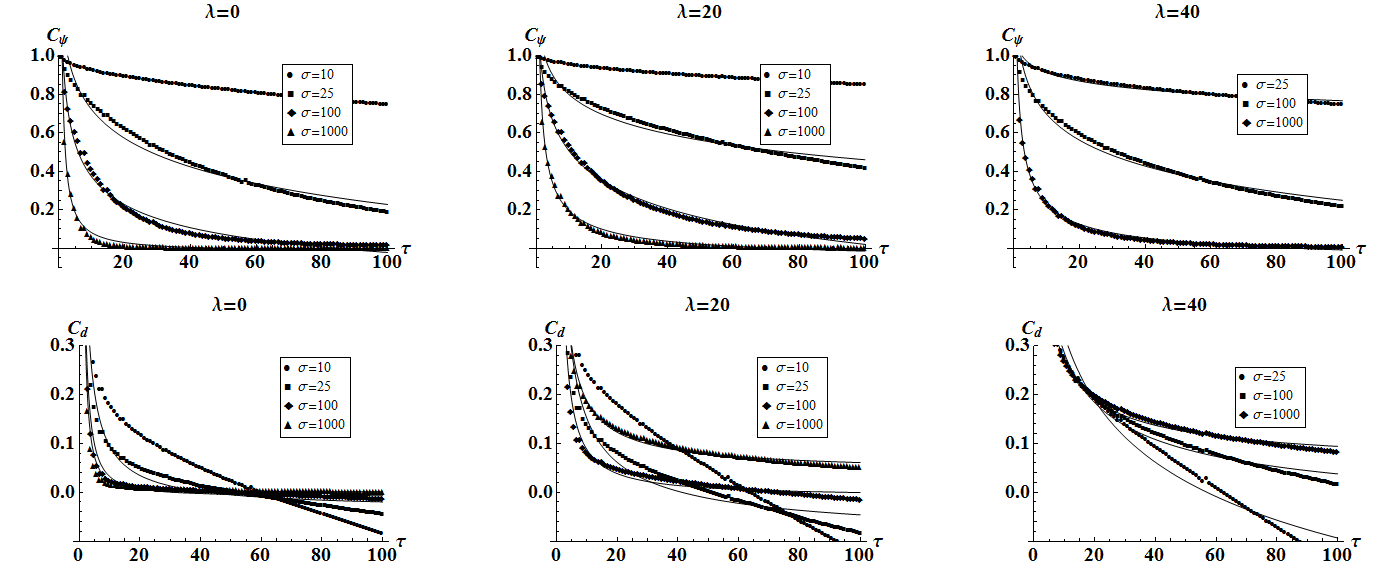}
\caption{A representative selection of the module length correlation functions $C_d(\tau)$ and the correlation functions of the angles between neighboring modules $C_\psi(\tau)$. Profiles has been fitted with function $a\tau^\alpha+c$.\label{pic:corr1}}
\end{figure*}

An example of collected data is presented in Figure \ref{pic:corr1}. In the high temperature regime (approximately for $\sigma>200$), both $C_\psi(\tau)$ and $C_d(\tau)$ are positive functions, asymptotically falling from 1 to 0, typical of the stochastic motion. However, they differ significantly in the low temperature regime. While $C_\psi(\tau)$ preserves its high temperature profile (but with values much closer to 1), the $C_d(\tau)$ reassembles a linear function, falling below 0 with the increase in $\tau$ . This long-term behavior of the low-temperature $C_d(\tau)$ indicates the domination of the deterministic motion in this temperature regime. In the context of the energetic landscape, introduced in section \ref{sec:pol_model}, we can suppose that the beads are trapped at the bottom of their potential energy wells and perform the damped oscillations, slightly perturbed by the noise. Apparently, while the beads' motion makes $d_j$ oscillate, it barely affects the angles, so the values of $C_\psi(\tau)$ are relatively close to 1. Additionally, the comparison between $C_\psi(\tau)$ and $C_d(\tau)$ suggest that the module length behavior evolves from deterministic into purely stochastic one as the temperature grows, while the $\psi_j(t)$ is of the stochastic nature for every $\sigma$.

%figure
\begin{figure}
\includegraphics[width=0.5\textwidth]{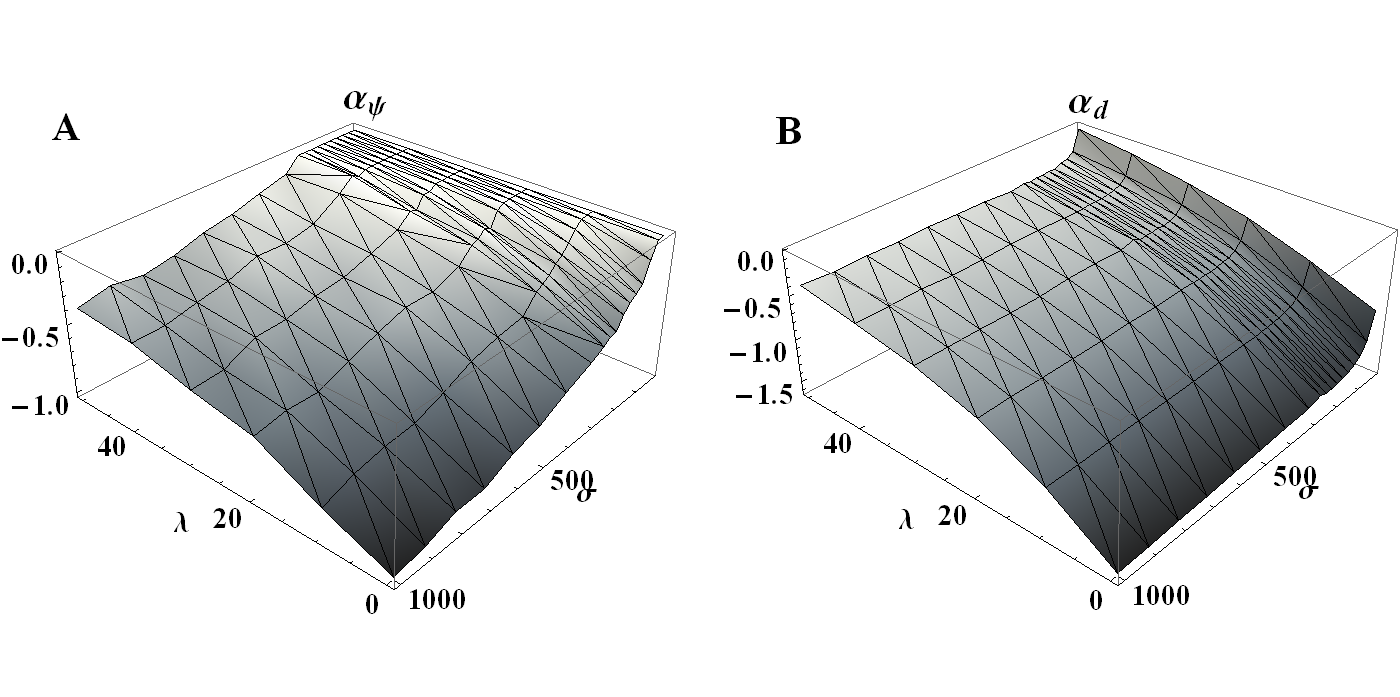}
\caption{The exponents $\alpha_d$ (A) and $\alpha_\psi$ (B) resulting from fitting the power function to correlation profiles $C_d(\tau)$ and $C_\psi(\tau)$. \label{pic:corr2} }
\end{figure}
 In order to measure the influence of $\sigma$ and the noise correlation length $\lambda$ on the $C_\psi(\tau)$ and $C_d(\tau)$, we have fitted the profile functions with the following model:
\begin{equation}
C(\tau)=a\tau^{\alpha}+c
\end{equation}
Despite inaccuracy for $\tau\to0$ and a divergence in the low-temperature regime, the power function model provides a quantitative information on $\sigma$ and $\lambda$ dependencies, thanks to the $\alpha$ parameter. The values of $\alpha_d$ and $\alpha_\psi$ plotted against the $\sigma$ and $\lambda$ are shown in Figure \ref{pic:corr2}. As expected, for all values of $\sigma$ and $\lambda$, $\alpha$ is negative, and tends to 0 with the decrease in temperature. However, for constant $\lambda$, $\alpha_\psi$ decreases at a similar pace with the growth of $\sigma$, while $\alpha_d$ varies slowly for the most of the temperature range, but jumps rapidly below $\sigma=100$.

The increase in the noise correlation length $\lambda$ affects both $\alpha_d$ and $\alpha_\psi$ in a similar way, namely, the larger $\lambda$, the lower $|\alpha|$ is obtained. This means that the correlation functions $C_\psi(\tau)$ and $C_d(\tau)$ decrease at slower rate and so the $d_j(t)$ and $\psi_j(t)$ vary less rapidly over time. Therefore, the dynamics of chain's shape becomes attenuated and a current conformation is preserved longer. 

%-----------------------AVERAGE MODULE LENGTH-----------------
\section{Module length distribution}
We have also investigated the marginal distribution $\Gamma(d)$ of the module length $d$ and its temperature evolution, with and without spatial correlations in noise. Taking into account that $d_j(t)$ may express an oscillatory behavior, we have reduced the time interval between data acquisitions to 1/4 of time unit, to avoid synchronization effects. The spatial resolution of histograms has been set to 0.21 length unit.  

\begin{figure}
\includegraphics[width=0.5\textwidth]{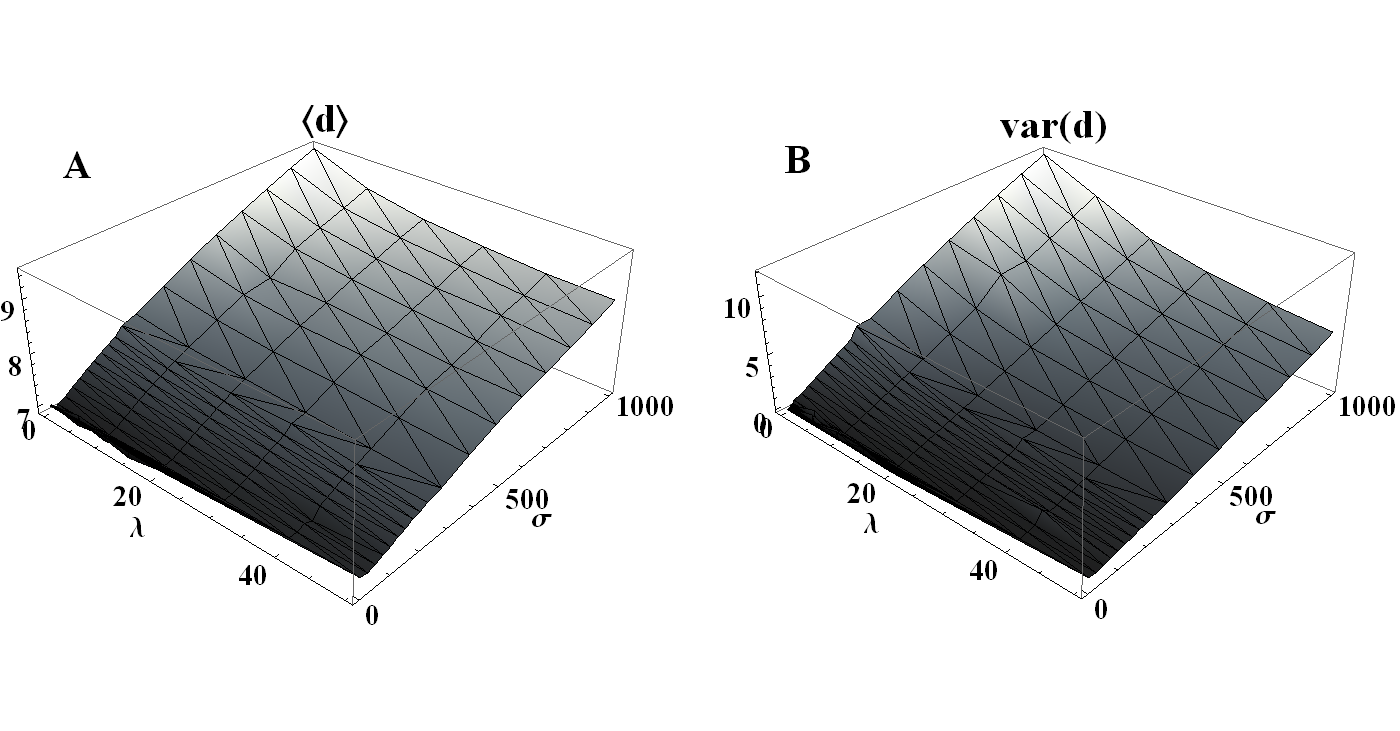}
\caption{Statistical characteristics of the module length distribution $\Gamma(d)$ as the functions of the noise correlation length $\lambda$ and the temperature $\sigma$: A) the mean module length $\left<d\right>$, B) the variance of $\Gamma(d)$. \label{pic:d_dist}}
\end{figure}

The profile of $\Gamma(d)$ proved to be a single peaked distribution, concentrated in the vicinity of its mean, with slight, but noticeable asymmetry. Therefore, in order to describe $\Gamma(d)$ we have calculated its first moment, second central moment (presented in Figure \ref{pic:d_dist}) and the skewness.
%figure

The skewness grows with $\sigma$ from 0 to approximately 0.6 and saturates at this value. Fortunately, asymmetry proved to be small enough, so the others parameters are still physically meaningful. The dispersion of distribution $\Gamma(d)$ grows with the increasing $\sigma$ (Figure \ref{pic:d_dist}B), which is expected diffusive behavior, but also the mean distance $\left< d\right>$ tends to grow (Figure \ref{pic:d_dist}A), starting from $\left< d\right>\approx d_0$. This fact, along with the non-zero skewness, indicates that the underlying potential is asymmetric, and, indeed, the presence of the repulsive Lennard-Jones core provides the reflective barrier preventing two neighboring beads from closing up. Conversely, the lengthening of $d_j$ is still possible as the energy well is not so steep for $d_j>d_0$ as in the opposite situation.

The spatial correlations  in noise play an inhibitory role for the process of the temperature dependent broadening of $\Gamma(d)$. $\lambda\neq0$ slows down the growth of both $\left< d\right>$ and the dispersion of $\Gamma(d)$. This effect can be explained by the following reasoning. When thermal bath imposes non-correlated, stochastic forces on beads $j$ and $j+1$, this commonly results in a nonzero relative force stretching (or shrinking) the module. However, when the noise is spatially correlated, stochastic forces applied to beads become similar at the length-scale of $\lambda$, which significantly reduces the relative forcing and, in turn, the $d$ is less affected by the noise. 

%-----------------------POLYMER UNFOLDING-----------------------
\section{Polymer unfolding}\label{sec:unfolding}
\begin{sidewaysfigure}
\includegraphics[width=\textwidth]{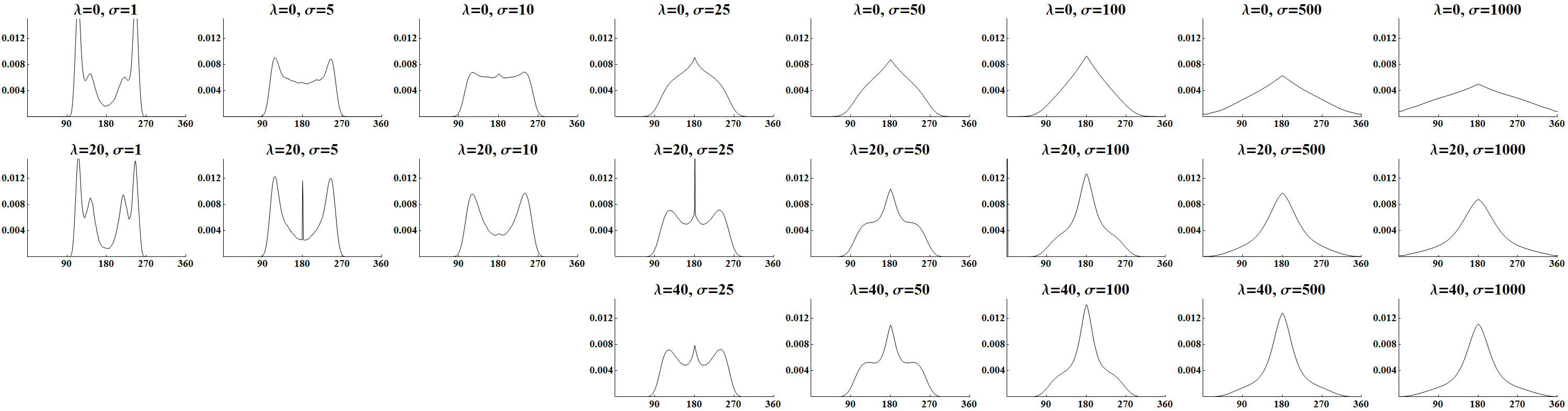}
\caption{The temperature evolution of angles distribution $\Phi(\psi)$ for noise correlation length $\lambda=0$, $\lambda=20$ and $\lambda=40$.\label{pic:unfold1}}
\end{sidewaysfigure}

The most unexpected effect that stems from the presence of the spatial correlations in the noise is the spontaneous linearization of the chain. Having defined the angles $\psi_j(t)$ in \eqref{eq:psi_j} we have been able to obtain a marginal distribution of angles $\Phi(\psi)$ depending on temperature $\sigma$ and correlation length $\lambda$. Similarly to previous section, the data has been collected every 1/4 of time unit, with the resolution of histogram set to $1^\circ$. The representative selection covering the entire range of tested parameters is presented in Figure \ref{pic:unfold1}. 

The temperature evolution of distribution $\Phi(\psi)$ gives an insight into how the angular degrees of freedom are freed with the rise in temperature. Let us analyze the $\lambda=0$ case, first. For low temperatures ($\sigma<10$) we obtain a symmetric bimodal distribution, which is in accordance with the predictions of the double minimum energetic landscape.  However, for an extremely low temperatures ($\sigma<3$) one can see 4 distinct peaks, which indicates that, probably, there are 2 additional minima. We can suppose that they are shallow, as they disappear fast with the rise in $\sigma$. For temperatures from $\sigma=5$ to $\sigma=13$ the increased penetration of the energy barrier region is viewable, and the third peak appears exactly at $\psi=180^\circ$. At $\sigma=13$ the two peaks indicating energy minimums can be no longer distinguished and from now on, the shape of distribution gradually transforms from a bell-like curve into a triangle profile, which is completed approximately for $\sigma=100$. Since then, the distribution broadens systematically with the increase in temperature. 

%figure
\begin{figure}
\includegraphics[width=0.5\textwidth]{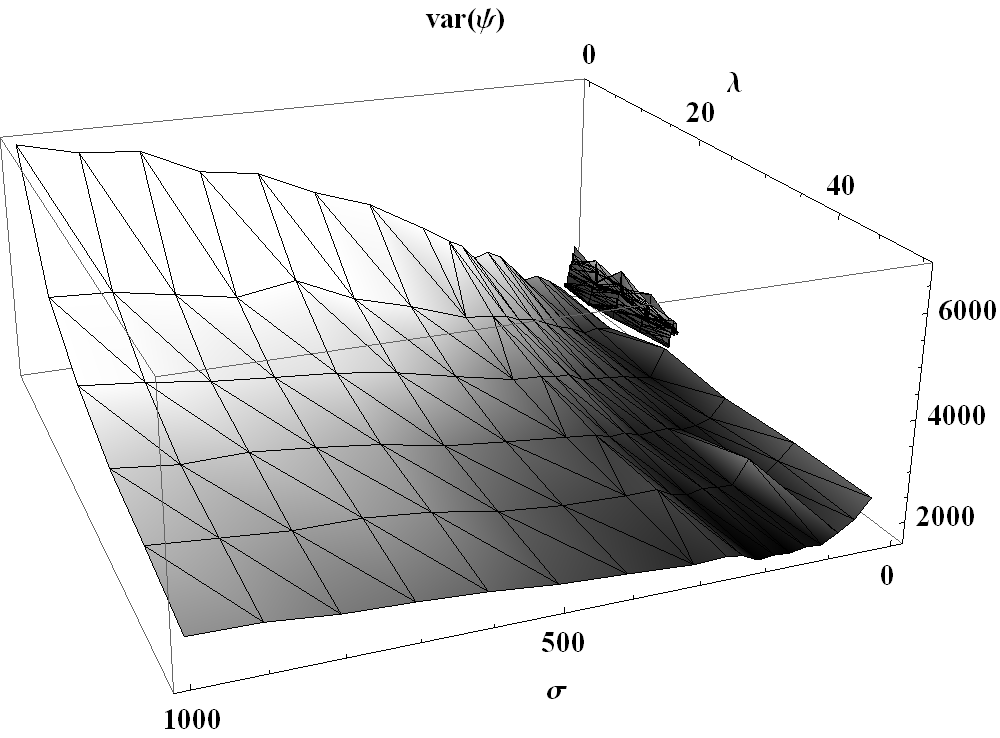}
\caption{The dispersion of distribution $\Phi(\psi)$ as a function of the noise correlation length $\lambda$ and the temperature $\sigma$.\label{pic:unfold2}}
\end{figure}

The introduction of $\lambda\neq0$ affects $\Phi(\psi)$ in a subtle, but remarkable way. In the low noise regime ($\sigma<25$) the increase in $\lambda$ retards the temperature evolution of $\Phi(\psi)$, so the bimodal profile is preserved in a wider interval of $\sigma$. However, when $\sigma$ exceeds 50, the profile transforms into a heavy-tailed peaked distribution, much more concentrated in the vicinity of $\psi=180^\circ$ than in the case of $\lambda=0$. The dispersion of $\Phi(\psi)$ is a measure of this effect, which shows that the higher $\lambda$, the lower value of the $\Phi(\psi)$'s second central moment. This indicates the linearization of chain, and thus, its unfolding. It is illustrated in Figure \ref{pic:unfold2}.

%figure

The single-peaked distribution with $\left< \psi \right>=180^\circ$ indirectly suggests that for high temperatures the system selects the single-minimum energetic configuration similar to the one shown in Figure \ref{pic:landscape}B rather than a double-minimum landscape in Figure \ref{pic:landscape}A. Knowing that the growth of the distance $l_j=|\vec{r}_{j+1}-\vec{r}_{j-1}|$ is a crucial factor in the merging of energetic minima, we can suppose that $l_j$ is a subject to a similar interplay as $d_j$, namely, the repulsive cores unable beads $j-1$ and $j+1$ to close up, while the stretching of $l_j$ is still possible. With the sufficiently high noise, this asymmetry could lead to the rise in $\left<l\right>$ and the domination of the single-minimum potential topology. This transition seems inherent to the system and it is present regardless of the spatial correlations in noise. Nevertheless, once the single-minimum state prevails, the bead can explore the well, provided there is enough relative forcing, and for $\lambda\neq0$ this forcing drops dramatically. In result, despite high noise amplitude, the bead is trapped near the minimum, so the angles between modules cannot vary as much as in the non-correlated case. This leads to narrowing of  $\Phi(\psi)$ around $\psi_j=180^\circ$. 

%-----------------SUMMARIZATION----------------------------------
\section{Summary}\label{sec:summary}
Summarizing our research, the most salient conclusion one can draw is that spatial correlations in thermal noise have an overall inhibitory effect on the system. This manifests in the general attenuation of the chain geometry dynamics both in the time domain, where the polymer tends to preserve its current shape, and in the temperature domain, where the evolution of statistical chain properties is retarded. It is also in agreement with our previous findings that the presence of nonzero correlations reduced the ability of chain to transfer between different conformations \cite{bib:mm}.

Such a behavior is not a surprise, as we can perceive the introduction of the spatial correlations into the thermal bath as a freezing of environment, and, in the limit $\lambda\to+\infty$, this should also lead to the complete attenuation of the system dynamics. In this context,  it is not solely the temperature of thermal bath that influences the system behavior, but also the structure of environment. Our approach, which decouples the temperature from the environmental correlation length, allows for more plasticity than the explicit simulation of thermal bath particles, yet requires proper scaling to avoid unphysical situations.


\begin{thebibliography}{99}
\bibitem{bib:sagues} F. Sagu\'{e}s, J. M. Sancho, J. Garc\'{\i}a-Ojalvo, \emph{Rev. Mod. Phys.}, \textbf{79}, 829 (2007)
\bibitem{bib:longa} R. Morgado, M. Cie\'{s}la, L. Longa, F. A. Oliveira, \emph{EPL}, \textbf{79}, 10002 (2007)
\bibitem{bib:gardiner} C. W. Gradiner, \emph{A handbook of stochastic processes}, 3rd ed. (Springer, Berlin, 2004)
\bibitem{bib:kou} S.C. Kou, \emph{Ann. Appl. Stat.}, vol. 2, \textbf{2}, 501-535 (2008)
\bibitem{bib:sed1} P. N. Segr\'{e}, E. Herbolzheimer, P. M. Chaikin, \emph{Phys. Rev. Lett.}, vol. 79, 13 (1997)
\bibitem{bib:sed2} E. Guazzelli, \emph{Phys. Fluids} 13, 1537 (2001)
\bibitem{bib:swimmers} P. T. Underhill, J. P. Hernandez-Ortiz, M. D. Graham, \emph{Phys. Rev. Lett.}, 100, 248101 (2008)
\bibitem{bib:binney} J. J. Binney et al., \emph{The theory of critical phenomena} (Oxford Univ. Pr., 1992)
\bibitem{bib:mosayebi} M. Mosayebi, E. Del Gado, P. Ilg, H. C. \"{O}ttinger \emph{Phys. Rev. Lett.}, 104, 205704 (2010)
\bibitem{bib:donati} C. Donati, S. C. Glotzer, P. H. Poole, W. Kob, S. J. Plimpton, \emph{Phys. Rev. E}, vol. 60, \textbf{3}, 3107 (1999)
\bibitem{bib:doliwa} B. Doliwa, A. Heuer, \emph{Phys. Rev. E}, vol. 61, 6 (2000)
\bibitem{bib:mitus} A. C. Mitus, A. Z. Patashinski, A. Patrykiejew, A. Sokolowski, \emph{Phys. Rev. B}, \textbf{66}, 184202 (2002)
\bibitem{bib:mm} M. Majka, P. F. G\'{o}ra, \emph{Acta Phys. Pol. B}, vol. 43, 1133 (2012)
\bibitem{bib:calorimetry} E. Donth, H. Huth, M. Beiner, \emph{J. Phys.: Condens. Matter}, 13, L451 (2001)
\bibitem{bib:impendspec} C. Dalle-Ferrier, C. Thibierge, C. Alba-Simionesco, L. Berthier, G. Biroli, J. P. Bouchaud, F. Ladieu, D. L'H\^{o}te, G. Tarjus, \emph{Phys. Rev. E}, 76, 041510 (2007)
\bibitem{bib:confocal} E. R. Weeks, J. C. Crocker, D. A. Weitz \emph{J. Phys.: Condens. Matter}, 19, 205131 (2007)
\bibitem{bib:kubo} R. Kubo, M. Toda, N. Hashitsume \emph{Statistical Mechanics II} (Springer, Berlin, 1985)
\bibitem{bib:golub} G. H. Golub, C. F. Van Loan, \emph{Matrix computations} (The Johns Hopkins Univ. Pr., 1996)
\bibitem{bib:wieczorkowski} R. Wieczorkowski, R. Zieli\'{n}ski, \emph{Komputerowe generatory liczb losowych}( WNT, Warsaw, 1997)
\bibitem{bib:kampen} N. G. van Kampen, \emph{Stochastic Processes in Physics and Chemistry}, (Elsavier Science Publishers, Amsterdam, 1987)
\bibitem{bib:kloeden} P. E. Kloeden, E. Platen \emph{Numerical Solutions of Stochastic Differential Equations} 3rd ed. (Springer-Verlag, Berlin, 1999)
\end{thebibliography}
\end{document}